\documentclass[authoryear, preprint,1p,times]{elsarticle}
\usepackage[table,xcdraw]{xcolor}
\usepackage{graphicx}
\usepackage{svg}
\usepackage{siunitx}
\usepackage{subcaption}
\usepackage{caption}
\usepackage{booktabs}
\usepackage{amsmath}
\usepackage{lineno}
\usepackage{ragged2e}
\usepackage{geometry}

\journal{}

\renewcommand{\boldsymbol}[1]{#1}
\renewcommand{\mathrm}[1]{#1}
\newcommand{\reals}{{\mbox{\bf R}}}

\begin{document}

\begin{frontmatter}

\title{Comparative analysis of whole-body center-of-mass estimation methods in dynamic and static activities using marker-based systems}

\author[1]{Jingshu Peng\fnref{cofirst}}

\author[2]{Mohsen Alizadeh Noghani\corref{cor1}\fnref{cofirst}}

\author[2]{Edgar Bolívar-Nieto}

\affiliation[1]{organization={Xingjian College, Tsinghua University},
            city={Beijing},
            postcode={100084}, 
            country={China}}

\affiliation[2]{organization={Aerospace and Mechanical Engineering Department, University of Notre Dame},
            city={Notre Dame},
            postcode={46556}, 
            state={IN},
            country={U.S.}}

\cortext[cor1]{Corresponding author. Email: malizade@nd.edu.}
\fntext[cofirst]{These authors contributed equally to this work and share first authorship.}

\begin{abstract}

Accurate estimation of the whole-body center of mass (CoM) is essential for assessing human stability and postural control. However, selecting the most accurate estimation method is challenging due to the complexity of human movement, diverse nature of activities, and varying availability of equipment, such as marker-based systems and ground reaction force (GRF) sensors. This study compares three CoM estimation methods -- ``Pelvis Markerset", ``Whole-Body Markerset", and ``Whole-Body Markerset \& GRFs" -- across static activities, such as standing with eyes closed, and dynamic activities, such as picking up an object from the ground. Using the root mean square (RMS) of ``external force residual" (the difference between measured ground reaction forces and estimated CoM accelerations multiplied by total body mass) as a performance metric, we found that while all methods performed similarly under static conditions, the ``Pelvis Markerset" method showed 96\% to 104\% higher RMS external force residual values during dynamic activities compared to the two whole-body methods ($p<0.001$, Cohen's $d$:2.90-3.04). The accuracy of ``Whole-Body Markerset \& GRFs" (i.e., Kalman filter) was similar to ``Whole-Body Markerset", suggesting that incorporating the GRFs through the presented Kalman filter does not improve the estimates from whole-body kinematics. Based on these findings, we recommend the ``Whole-Body Markerset" as it performs well and does not require information from GRFs. The ``Pelvis Markerset" method can be used in static activities or when markersets around the pelvis reflect whole-body kinematics. This method is not recommended for CoM state estimation in highly dynamic scenarios and when whole-body markersets are available.

\end{abstract}

\end{frontmatter}

\section{Introduction}

The whole-body center of mass (CoM) is a crucial parameter in the human postural control system and plays a significant role in assessing the stability of human motion \citep{hof2005conditiona,corriveau2001postural,schepers2009ambulatory}. Stability in static conditions is maintained when the vertical projection of the CoM remains within the base of support (BoS) \citep{winter1995human}. In dynamic conditions (e.g., walking), stability depends not only on the whole-body CoM position but also on its velocity \citep{pai1997centera}. The ``extrapolated center of mass" (XCoM) integrates both, predicting the CoM’s future position. Stability is maintained if the XCoM stays within the BoS, providing a margin of stability that helps prevent falls \citep{hof2005conditiona,hof2008extrapolateda}. These principles also apply to legged robots \citep{sangjoo2007estimation}, where precise CoM control is crucial for maintaining balance. As a result, whole-body CoM estimation is widely used in robotics and human movement research \citep{cotton2009estimation,catena2018comparison,buurke2023comparisona}, particularly in wearable robotics, where it supports balance control strategies \citep{rajasekaran2015adaptive} and motion planning \citep{kagawa2015optimization}.

There are two main approaches for estimating whole-body CoM motion, each with its own underlying assumptions. One approach, based on \emph{double integration} of ground reaction forces (GRFs), calculates the CoM position given external forces measured by force plates or sensors \citep{donelan2002simultaneous,schepers2009ambulatory}. The second method, referred to as \emph{segment kinematics}, models the body as a chain of rigid segments and estimates the overall CoM by calculating the kinematics of the CoM for one or multiple segments. \citep{hasan1996simultaneous,pavol2002body,schepers2009ambulatory}. Within \emph{segment kinematics}, three main techniques are employed to estimate the CoM position: (1) using the position of one or more markers on the pelvis \citep{saini1998vertical}, (2) employing a whole-body marker set with anthropomorphic tables to estimate the segment CoM and mass, which can be used to calculate the whole-body CoM \citep{delp2007opensima, dumas2007adjustmentsa, winter2009biomechanics}, and (3) combining the whole-body marker set technique with GRF measurements \citep{xinjilefu2015center}. Due to their popularity and high accessibility of sensor data in contemporary research, we will focus on these three methods, hereon referred to as ``Pelvis Markerset", ``Whole-Body Markerset", and ``Whole-Body Markerset \& GRFs".

The primary goal of this study is to compare the effectiveness of the three methods in estimating whole-body CoM position and velocity during 14 distinct activities, including static activities (e.g., standing with eyes closed) and dynamic activities (e.g., picking up an object from the ground). Section 2 outlines the three CoM estimation methods and introduces the ``external force residual" as the key performance metric. The root mean square (RMS) of this residual is used to quantify the differences in accuracy across methods. This comprehensive analysis of various activities provides valuable insight into the most suitable estimation method, considering movement complexity and available instrumentation. Section 3 presents the results of CoM position and velocity estimations, while Section 4 offers comparisons and recommendations based on these results. Finally, Section 5 summarizes the key findings.

\section{Methods}
 \subsection{Experimental procedure and data collection}

We used an existing data set \citep{alizadeh2024predicting} of ten healthy subjects (mean $\pm$ SD, age: 22.9 $\pm$ 3.5 years, height: 1.73 $\pm$ 0.083 m, mass: 63.66 $\pm$ 9.79 kg, 5 males) who provided written informed consent to participate in the study approved by the University of Notre Dame Institutional Review Board (Protocol ID: 24-01-8277).

The subjects were fitted with fifty-seven reflective markers following a whole-body marker set \citep{optitrack_biomech57} (Fig.~\ref{fig:1}). Then, they performed a series of 14 activities, each repeated three times, based on the Berg Balance Scale test \citep{berg1992measuring} over a walkway instrumented with force plates (Table~\ref{tab:activity_list}). Based on the magnitude of displacement of body segments, the activities were qualitatively classified as Static or Dynamic.

\begin{table}[t!]
\centering
\caption{The list of activities performed by participants and the provided instructions. The activities marked with {\textdagger} are in the Static group, while others are considered Dynamic. Shaded activities are similar to those in the Berg Balance Scale test. This table has been reproduced from our previous work \citep{alizadeh2024predicting}.}
\label{tab:activity_list}
\small
\begin{tabular}{p{3.2cm}p{8cm}}
\toprule
\textbf{Activity} & \textbf{Description} \\ \midrule
\rowcolor{gray!10}
(1) Sit to Stand & Stand up from the chair without using your hands for support. \\  
\rowcolor{gray!10}
(2) Stand to Sit & Sit down on the chair without using your hands for support.\\
\rowcolor{gray!10}
(3) Foot on Chair & Place your right foot on the upper surface of the chair in front of it. The left foot replicates this motion but is not placed on the chair. Repeat for four cycles. \\ 
(4) Stride & Take a stride forward, then stop. \\ 
\rowcolor{gray!10}
(5) Look Behind\textsuperscript{\textdagger} & Turn to look directly behind, first over your left shoulder, and then your right shoulder.  \\ 
\rowcolor{gray!10}
(6) Feet Together\textsuperscript{\textdagger} & Stand with your feet placed as close together as possible. \\ 
\rowcolor{gray!10}
(7) Eyes Closed\textsuperscript{\textdagger} & Stand still with your eyes closed. \\ 
\rowcolor{gray!10}
(8) Turn & Turn 180°, pause, then turn back to your original direction. \\ 
\rowcolor{gray!10}
(9) Tandem Feet\textsuperscript{\textdagger} & Stand with one foot directly in front of the other, with your feet as close together as possible. \\ 
\rowcolor{gray!10}
(10) One Leg & Stand on one leg, then return to neutral pose. \\ 
(11) Squat & Perform a squat, then return to neutral pose.\\ 
\rowcolor{gray!10}
(12) Reach Forward & Reach forward as much as possible while keeping your arms parallel, then return to neutral pose.  \\ 
\rowcolor{gray!10}
(13) Object Pickup & Pick up an object placed in front of your right foot. \\ 
(14) Shoe Lace & Step down to emulate tying your right shoe laces, then return to neutral pose. \\ \bottomrule
\end{tabular}
\end{table}

During experiments, the force plates (Kistler Instrumente AG, Winterthur, Switzerland) recorded 3D GRFs at \SI{1000}{\hertz}, while a twelve-camera motion capture system (PrimeX22, NaturalPoint Inc., OR, USA) simultaneously captured the marker positions at \SI{200}{\hertz}. The marker data was processed using Motive 3.0 (NaturalPoint Inc., OR, USA) and lowpass-filtered at \SI{6}{\hertz}. Forceplate data was lowpass-filtered using a 5th order Butterworth filter (cutoff frequency of \SI{20}{\hertz}) and then downsampled to \SI{200}{\hertz}. Subsequent analysis was performed in MATLAB ( MathWorks Inc., MA, USA) and OpenSim \citep{delp2007opensima}. The One Leg, Squat, and Shoe Lace trials were split into two parts, capturing the movements at the start and return phases of the movement.

\subsection{CoM state estimation}

\begin{figure}
    \centering
    \begin{subfigure}{0.65\textwidth}
        \centering
        \includegraphics[width=\linewidth]{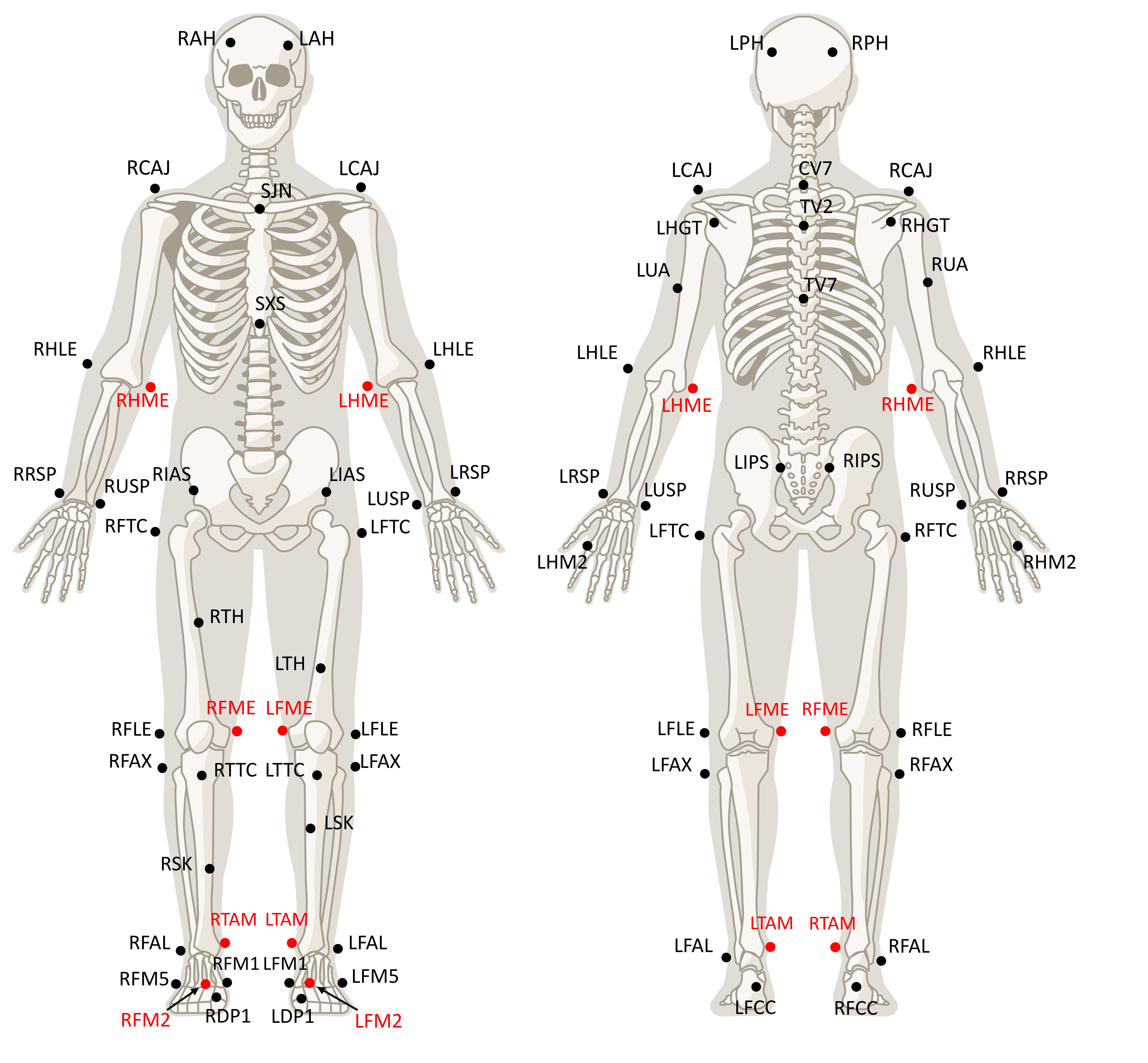}
        \caption{}
        \label{fig:markerset}
    \end{subfigure}\hfill
    \begin{subfigure}{0.35\textwidth}
        \centering
        \includegraphics[width=\linewidth]{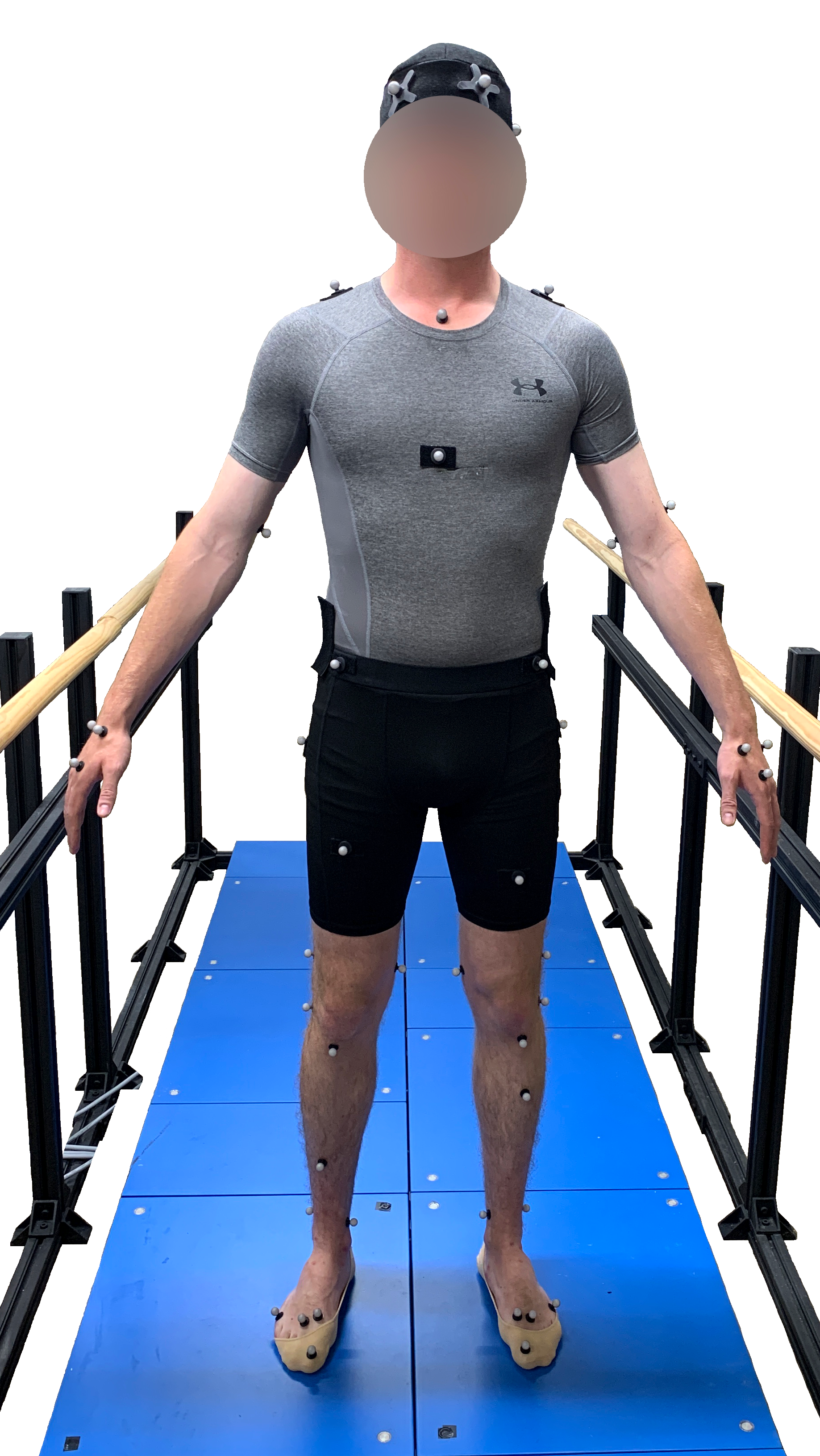}
        \caption{}
        \label{fig:subject}
    \end{subfigure}
    \caption{(a) The anterior and posterior views of the Biomech (57) markerset. The calibration markers are highlighted in red. (b) A participant fitted with the markerset, standing in an A-pose on the instrumented walkway.}
    \label{fig:1}
\end{figure}

We employed three methods to estimate the CoM position and velocity. For the Pelvis Markerset ($PM$) method, the CoM position was calculated by averaging the positions of four pelvic markers: RIAS, LIAS, LIPS, and RIPS (Fig.~\ref{fig:markerset}) \citep{saini1998vertical}. The CoM velocity was obtained by applying central finite differences to the CoM position. 

For the Whole-Body Markerset ($WM$) method, we modified a full-body OpenSim model \citep{rajagopal2016fulla} by adding a separate head-and-neck segment connected to the torso with three rotation degrees of freedom. The segment's mass and CoM position were derived from the anthropometric tables \citep{dumas2007adjustmentsa}. After performing inverse kinematics, the whole-body CoM position and velocity were obtained using OpenSim's BodyKinematics analysis tool \citep{seth2011opensima,sherman2011simbodya}.

Lastly, the Whole-Body Markerset \& GRFs ($WMG$) method fused the $WM$ estimates with GRFs to compute the CoM position and velocity \citep{xinjilefu2015center}. The dynamics of the CoM were modeled as follows,
\begin{gather}
    \boldsymbol{x}[k] = \mathrm{A} \boldsymbol{x}[k-1] + \mathrm{B} \boldsymbol{u}[k-1] + \boldsymbol{w}[k] \label{eq:disc_state}, \\
    \boldsymbol{y}[k] = \mathrm{C} \boldsymbol{x}[k] + \boldsymbol{v}[k],\\
            A = \begin{bmatrix}
1 & \Delta T & 0 & 0 & 0 & 0 \\
0 & 1 & 0 & 0 & 0 & 0 \\
0 & 0 & 1 & \Delta T & 0 & 0 \\
0 & 0 & 0 & 1 & 0 & 0 \\
0 & 0 & 0 & 0 & 1 & \Delta T \\
0 & 0 & 0 & 0 & 0 & 1 
\end{bmatrix}, B = 
\begin{bmatrix}
\frac{\Delta T^2}{2} & 0 & 0 \\
\Delta T & 0 & 0 \\
0 & \frac{\Delta T^2}{2} & 0 \\
0 & \Delta T & 0 \\
0 & 0 & \frac{\Delta T^2}{2} \\
0 & 0 & \Delta T
\end{bmatrix},
C = \begin{bmatrix}
1 & 0 & 0 & 0 & 0 & 0 \\
0 & 0 & 1 & 0 & 0 & 0 \\
0 & 0 & 0 & 0 & 1 & 0 
\end{bmatrix},
\end{gather}
where $\boldsymbol{x} \in \reals^6$ represents the 3D components of the CoM position (i.e., $x_1$, $x_3$, and $x_5$) and velocity (i.e., $x_2$, $x_4$, and $x_6$), which are to be estimated;  $\boldsymbol{u} \in \reals^3$ is the acceleration of the CoM computed from GRFs measured by the force plates; and $\boldsymbol{y} \in \reals^3$ is the nonsmoothed CoM position given by $WM$. 
%$\mathrm{A}$ and $\mathrm{B}$ correspond to double-integrator dynamics discretized with zero-order hold assumption at the sampling frequency of \SI{200}{\hertz}. 
$\mathrm{A}$ is the state transition dynamics, $\mathrm{B}$ is the influence of control inputs on the state, $\mathrm{C}$ is the mapping from the states to observed outputs (i.e., CoM position), and $\Delta T =$ \SI{5}{\milli\second} is the sample time. We assumed that the process and measurement noise ($\boldsymbol{w}[k] \sim \mathcal{N}(0, \mathrm{V_{w}})$ and $\boldsymbol{v}[k] \sim \mathcal{N}(0, \mathrm{V_{v}})$) follow a normal distribution with zero mean and process and measurement covariance matrices $V_w$ and $V_v$. We applied a Kalman filter to estimate $x[k]$. 
%A genetic algorithm that minimizes the ``external force residual" specified the process and measurement covariance matrices, $V_w$ and $V_v$.
A genetic algorithm that minimized the difference between the $WMG$ and low-pass-filtered $WM$ estimates (with a cutoff frequency of \SI{6}{\hertz}) specified the process and measurement covariance matrices, $V_w$ and $V_v$.

The accuracy of these methods in estimating CoM position and velocity was quantitatively evaluated using the ``external force residual", $r(t) \in \reals^3$, defined as
\begin{equation}
r(t) = F_{ext}(t) - ma_{est}(t),
\label{residual}
\end{equation}
where $F_{ext}(t)$ represents the net external force at time $t$, $m$ is the participant's mass, and $a_{est}$ is the estimated acceleration. This residual measures the discrepancy between the net external force, computed by subtracting gravitational force from the force plate measurements, and the estimated net external force based on the acceleration given by the three estimation methods, providing a metric for assessing their accuracy. We used the ``external force residual" because the experimental GRFs divided by total mass were the closest available measurement to a ``ground truth" for CoM kinematics.

To minimize the noise that was introduced during the estimation process, a smoothing technique minimized both the norm of the signal fit error and the third derivative of the signal \citep{yazdani2012simple}, as formulated by the following optimization problem,
\begin{equation}
\min_{\hat{x}} \left\| x - \hat{x} \right\|_2^2 + \alpha \left\| D_3 \hat{x} \right\|_2^2,
\label{smoothing}
\end{equation}
where $x$ is the original signal, $\hat{x}$ is the filtered signal, $\left\|.\right\|_2^2$ is the $L^2$-norm, and $D_3$ denotes the third derivative; therefore, $\left\| D_3 \hat{x} \right\|_2^2$ penalizes the CoM jerk when applied to the position signal, and CoM snap when applied to the velocity signal, with the tunable parameter $\alpha$ controlling the trade-off between signal fit and jerk or snap minimization \citep{boyd2004convex}. Through experimentation, a value of 300 for $\alpha$ was found to balance smoothing the signal while preserving local features. For all three methods, following the smoothing process, $a_{est}$ was obtained by applying central finite differences for numerical differentiation.

After computing the residual $r(t)$, we compared the performance of the methods by averaging the RMS of $r(t)$ along the three dimensions and subsequently the three repeats of each activity. The normality of the data was verified using Q-Q plots. One-way analysis of variance (ANOVA) was used to identify if the estimation method was a main effect and post-hoc t-tests with Bonferroni correction were performed to detect pairwise differences.  Significance level ($\alpha$) was set to 0.05 and eta-squared ($\eta^2$) and point estimate of Cohen's $d$, along with its 95\% confidence interval (95\% CI) quantified effect sizes; magnitudes $d$ of 0.2, 0.5, and 0.8 were considered as thresholds for small, medium, and large effects, respectively \citep{cohen2013statistical}. 
%Paragraph 4: Describe the statistical analysis methods used.

\section{Results}

\begin{figure}
    \centering
    % First row of subfigures
    \begin{subfigure}{0.48\textwidth}
        \centering
        \includegraphics[width=\linewidth]{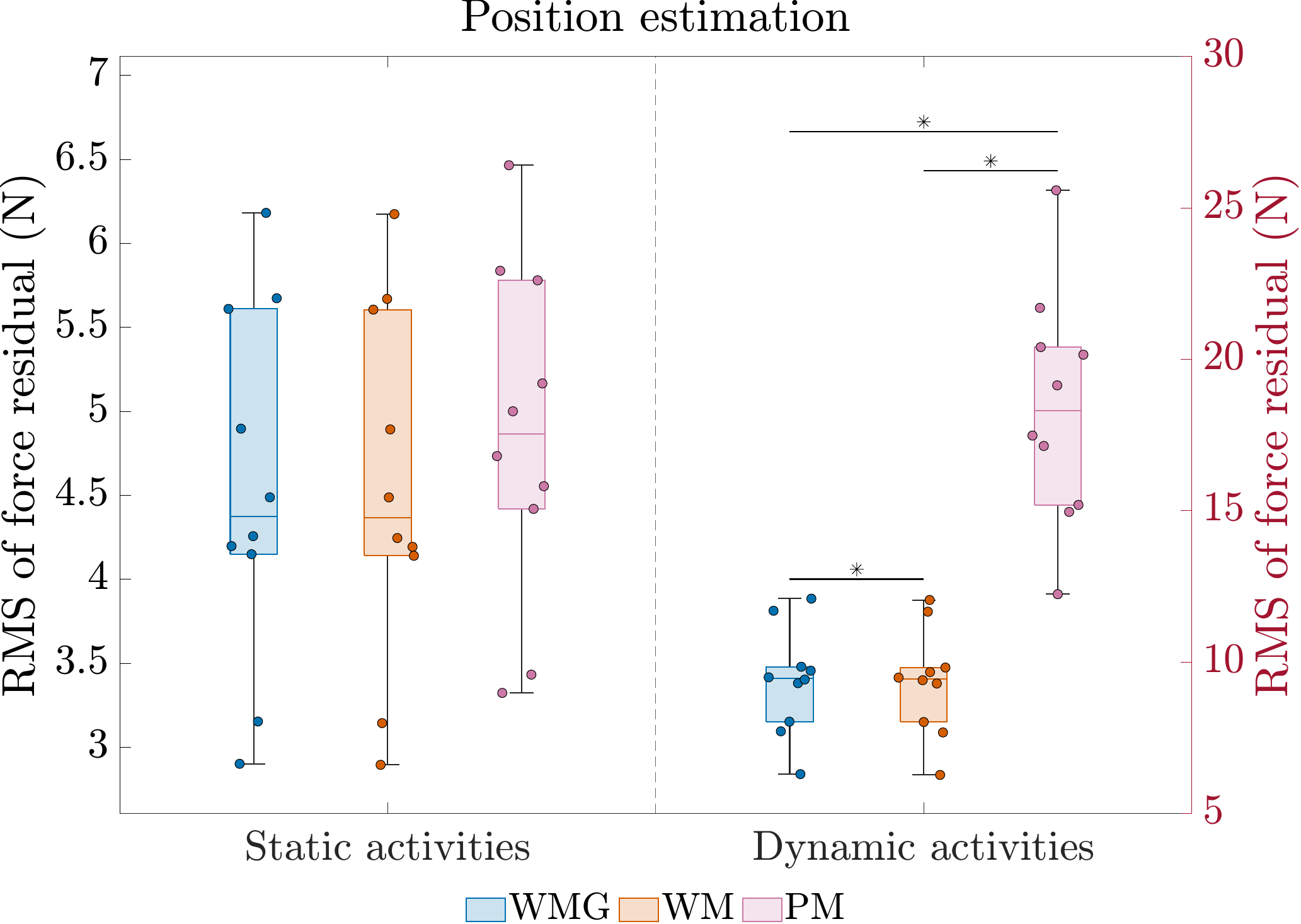}
        \caption{}
        \label{fig:pos_est_box}
    \end{subfigure}\hfill
    \begin{subfigure}{0.48\textwidth}
        \centering
        \includegraphics[width=\linewidth]{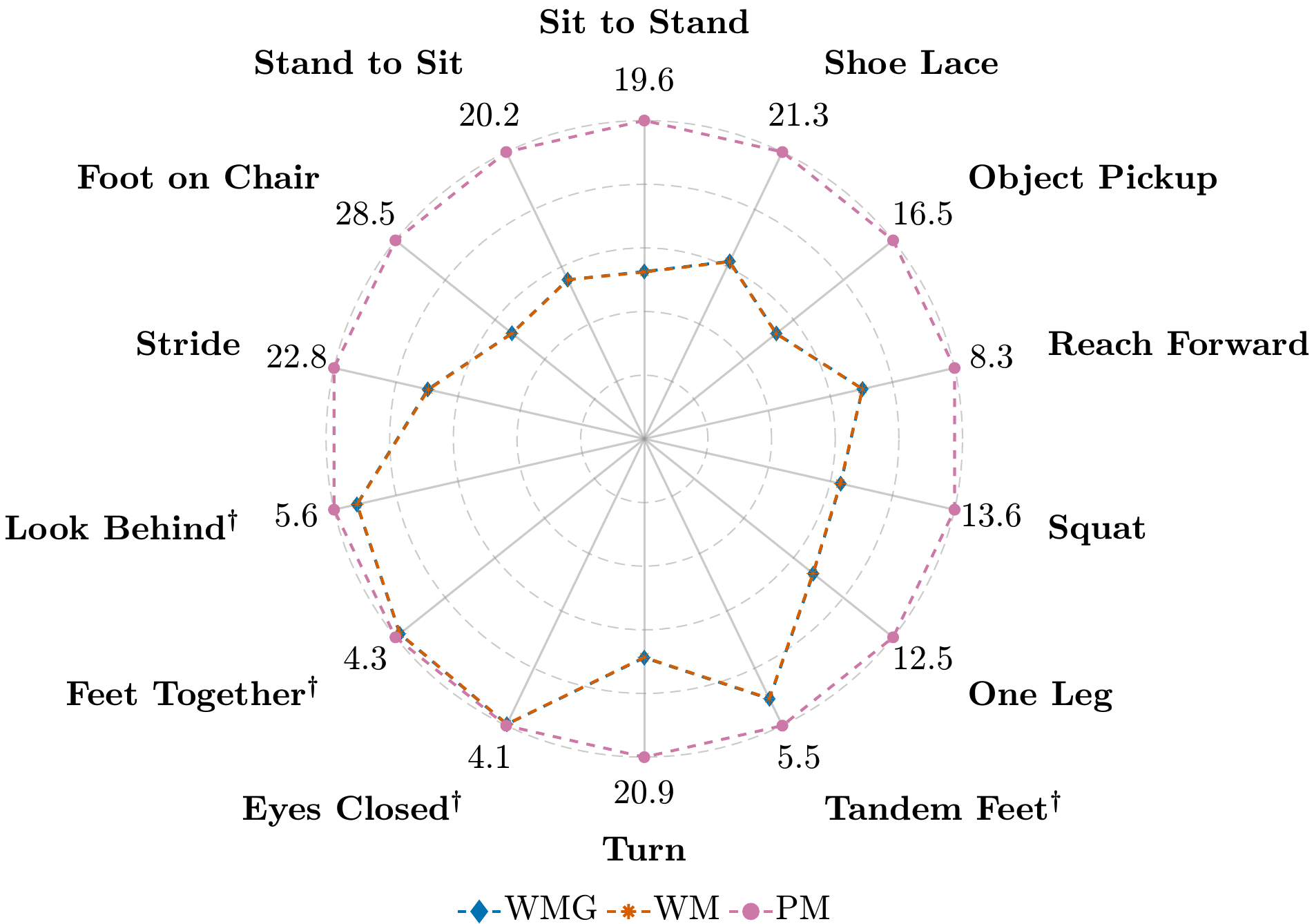}
        \caption{}
        \label{fig:pos_est_radar}
    \end{subfigure}
    
    \vspace{0.1cm} % Adjust space between rows
    
    % Second row of subfigures
    \begin{subfigure}{0.48\textwidth}
        \centering
        \includegraphics[width=\linewidth]{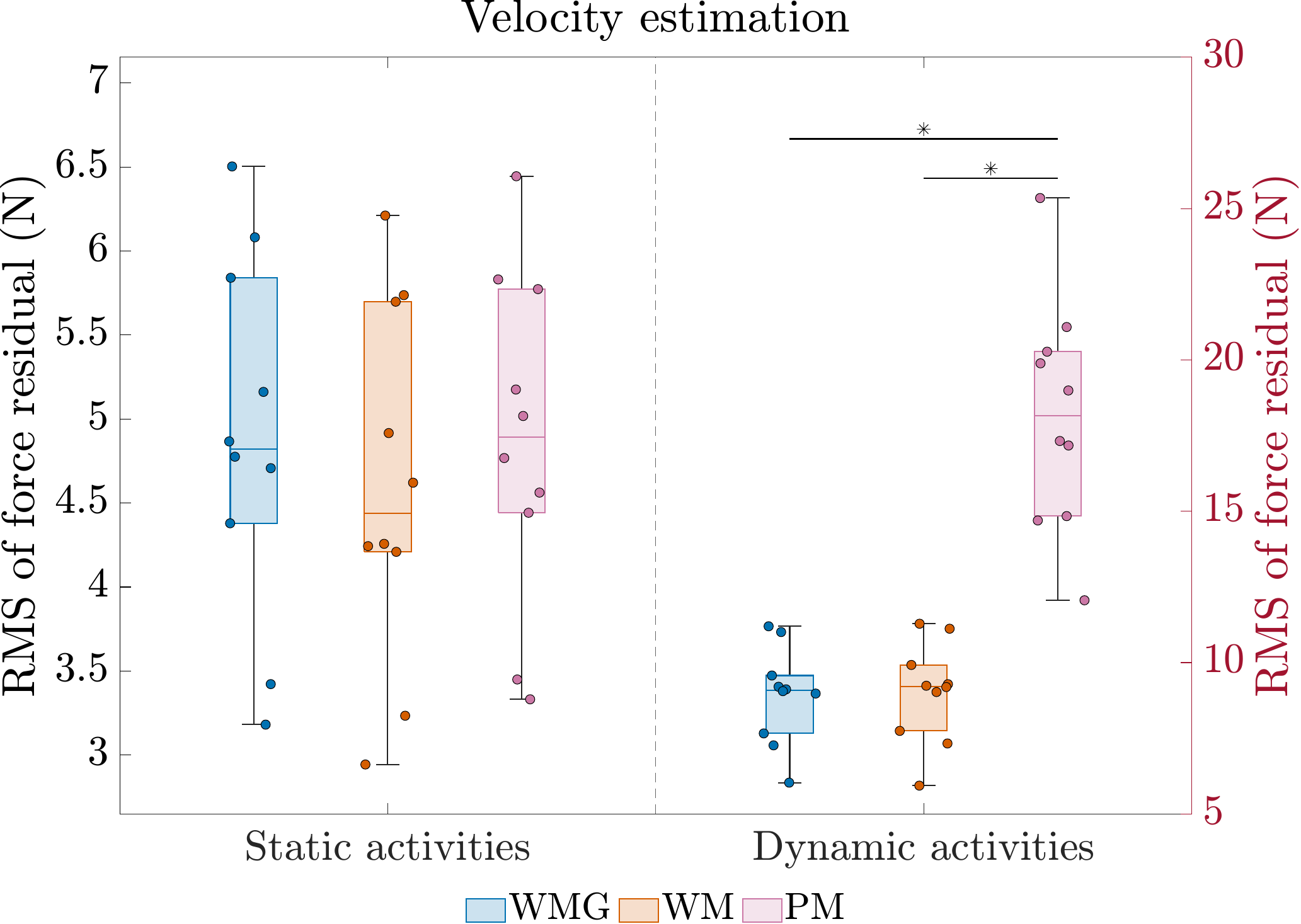} % Replace with the actual figure path
        \caption{}
        \label{fig:vel_est_box}
    \end{subfigure}\hfill
    \begin{subfigure}{0.48\textwidth}
        \centering
        \includegraphics[width=\linewidth]{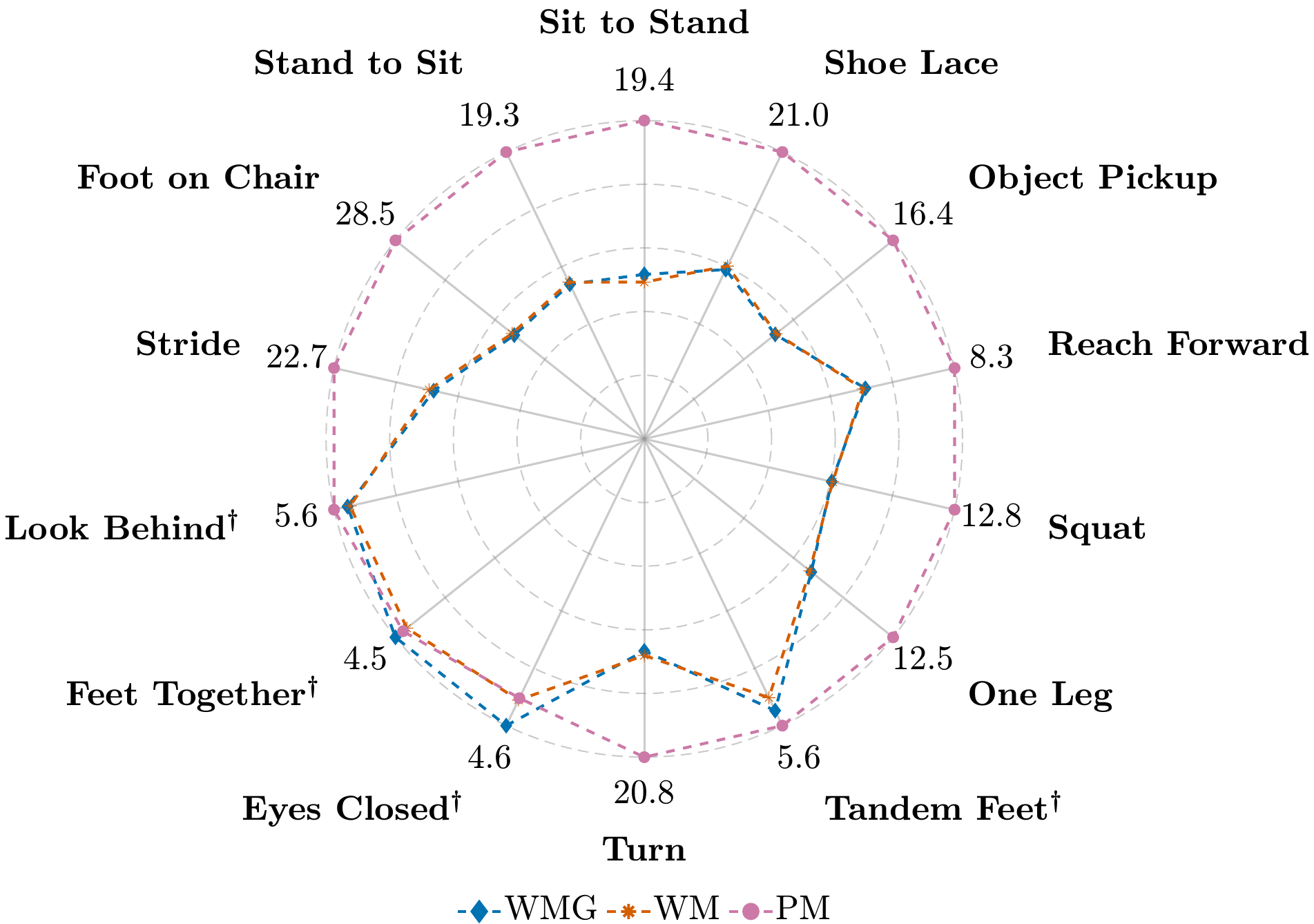} % Replace with the actual figure path
        \caption{}
        \label{fig:fig:vel_est_radar}
    \end{subfigure}
    
    \caption{(a,c) RMS of force residual for Static and Dynamic activities, when estimating position and velocity. The * symbols denote statistical significance in the post-hoc tests. (b,d) The average RMS of force residual for the 14 activities; the four activities in the Static group are denoted by \textdagger. The axis limits represent the maximum RMS residual among the methods in each activity.}
    \label{fig:overall}
\end{figure}

ANOVA did not identify a significant difference among the three methods when estimating position in the Static activities ($p=0.730$) (Fig.~\ref{fig:pos_est_box} and \ref{fig:pos_est_radar}). However, there was a main effect of estimation method in the Dynamic activities ($p<0.001$, $\eta^2$ = 0.74). Post-hoc tests identified significant differences between all three pairs of $WMG-WM$ ($p<0.001$, $d=0.01$, 95\% CI = $(-0.82, 0.85)$), $WMG-PM$ ($p<0.001$, $d=-2.90$, 95\% CI = $(-4.15, -1.63)$), and $WM-PM$ ($p<0.001$, $d=-2.91$, 95\% CI = $(-4.16, -1.63)$).

For CoM velocity, while there was no main effect of the estimation method in the Static group ($p=0.791$), the difference in the Dynamic activities was significant ($p<0.001$, $\eta^2=0.76$) (Fig.~\ref{fig:vel_est_box} and \ref{fig:fig:vel_est_radar}). Subsequent t-tests revealed differences between $WMG-PM$ ($p<0.001$, $d=3.04$, 95\% CI = $(-4.32, -1.73)$) and $WM-PM$ ($p<0.001$, $d=2.99$, 95\% CI = $(-4.33, -1.69)$) pairs, but not between $WMG-WM$ ($p=0.066$).

\section{Discussion}
\subsection{Comparison of the CoM estimation methods}

When estimating position in Static activities, there is no method among the three that outperforms the others. This was expected because in the four activities in the Static group (Look Behind, Feet Together, Eyes Closed, and Tandem Feet), the participants remained upright and the body segments did not move significantly. As a result, the CoM remained close to the center of pelvis, and the $PM$ method captured the CoM kinematics with similar accuracy to the other two methods. In contrast, in the Dynamic activities, $PM$ performed worse than the other two methods, with the average RMS residual being 96\% higher than $WM$ (Fig.~\ref{fig:pos_est_box}). This trend was consistent across all the dynamic activities (Fig.~\ref{fig:pos_est_radar}); the $PM$ method does not include information about the movement of the upper or lower body, which experienced large displacements in the Dynamic activities such as Shoe Lace, resulting in the large RMS values. Interestingly, the RMS of residual of $WMG$ was consistently and slightly worse than $WM$ across all activities, with a 0.28\% higher average. However, the effect size was negligible ($d=0.01$) and its 95\% CI covered zero, suggesting that this may not translate to a difference in the performance of the estimators. 

The results for velocity estimations were similar to position, albeit with no difference between the two whole-body methods in the Dynamic activities (Fig.~\ref{fig:vel_est_box}). We note that the $WMG$ had a larger RMS than the pelvis-based method in Eyes Closed and Feet Together activities (Fig.~\ref{fig:fig:vel_est_radar}). This may be due to the Kalman filter estimate converging to a velocity estimate with a small offset from 0 in these Static trials, resulting in increased RMS residual force in the process.

\subsection{Recommendations for the CoM estimation methods}
 
The results suggest that incorporating GRFs through the presented Kalman filtering framework does not produce a more accurate estimate over the $WM$ method. Therefore, to estimate the CoM position and velocity, the $WM$ method is recommended for its consistent accuracy across all activities. In resource-limited situations, the $PM$ can be a feasible option for static activities. However, its performance gap with the whole-body kinematics methods became more pronounced with increased CoM movement. In particular, when substantial CoM displacement was involved, this disparity was large (Fig.~\ref{fig:pos_est_radar} and \ref{fig:fig:vel_est_radar}). Therefore, caution is required when interpreting the CoM state estimation results in dynamic activities using the $PM$ method.

\section{Conclusion}

Our study compared three CoM position and velocity estimation methods across 14 distinct activities in 3D by assessing their accuracy through the RMS of ``external force residual". The results show the whole-body marker set delivers accurate estimates across all the activities, while those based on the pelvis marker set are mainly effective in static activities. Further, fusing the estimates from the whole-body marker set with the GRFs by using a Kalman filter did not improve the estimation. These findings across a diverse array of movements provide practical recommendations for selecting a CoM estimation method tailored to the dynamics of the movement.

\bibliographystyle{elsarticle-harv}
\bibliography{ref_updated}

\end{document}